\documentclass[11pt]{article}
\pdfoutput=1
\usepackage[utf8x]{inputenc}
\usepackage{amssymb,amsmath,mathrsfs,enumerate}
\usepackage{amsthm}
\usepackage{graphicx,rotate,multicol,braket, multirow}
\usepackage{url}
\usepackage{cite}
\usepackage{soul}
\usepackage{braket}
\usepackage[normalem]{ulem}
\usepackage{wrapfig}
\usepackage[textfont=it,labelfont=bf]{caption}
\usepackage{bm}
\usepackage[colorlinks=true,
linkcolor=red,
urlcolor=blue,
citecolor=blue]{hyperref}
\usepackage{lineno}
\usepackage{color}
\usepackage{bbold}
\usepackage[dvipsnames]{xcolor}
\long\def\rpl#1!!#2!!{\textcolor{red}{#1} \textcolor{blue}{#2}}

\usepackage[T1]{fontenc}

\def\bar{\overline}

\newcommand{\be}{\begin{equation}}
\newcommand{\Tbb}{\bar T_{\beta\beta}}
\newcommand{\ee}{\end{equation}}
\newcommand{\bea}{\begin{eqnarray}}
\newcommand{\eea}{\end{eqnarray}}

\evensidemargin=\oddsidemargin
\def\Eqn#1{Eq.\ (\ref{#1})}

\allowdisplaybreaks

\title{\Large\bf 
Towards a Proof of the Improved Quantum Null Energy Condition
}

\author{
  \sf 
  Ido Ben-Dayan,$^{a,}$\footnote{idobd@ariel.ac.il}
\quad Ayushi Srivastava$^{a,}$\footnote{srivastavaayushi860@gmail.com}
\\[10pt]
    \small\em$^a$Physics Department, Ariel University, Ariel 40700, Israel
}

\date{}

\begin{document}


\maketitle

\begin{abstract}
The Improved Quantum Null Energy Condition (INEC) was recently derived from the (restricted) quantum focusing conjecture (QFC), and is a statement about the energy-momentum tensor (EMT) of field theories in Minkowski space-time. It extends the quantum null energy condition (QNEC), and includes the possibility of expanding or contracting geodesics. 
Treating the INEC as a field theory inequality, we use the properties of relative entropy and modular Hamiltonian associated with null deformation of the sphere to recast it as an integrated inequality. At leading order in a perturbation of the vacuum, this inequality becomes a weighted positivity condition on the smeared null energy. As the underlying single-segment light-cone operator is unbounded below for $d>2$, we pass to the Lorentzian cylinder $\mathbb{R}\times S^{d-1}$, where the operator is well-defined. We verify that the INEC holds explicitly in the thermal state, in both the perturbative and high-temperature regimes. Furthermore, using the QNEC and INEC as a basis, we briefly speculate about a possible modified Quantum Focusing Conjecture. 
\end{abstract}

\bigskip

\section{Introduction} \label{s:intro}
Energy and its conservation is a cornerstone of Physics. For any well-defined physical system, we expect the energy or more precisely the Hamiltonian operator to be bounded from below to ensure a stable ground state. In the absence of gravity, it is a straightforward notion since only energy differences matter. In general relativity (GR) this is a more subtle statement because all forms of energy gravitate, and determine the dynamics of space-time, via Einstein's field equations (EFE), 
\begin{eqnarray}
\label{e:eeqn}
G_{\mu \nu}\equiv R_{\mu \nu }- \frac{1}{2}R g_{\mu \nu } = 8\pi G T_{\mu \nu}\,.
\end{eqnarray}
where $T_{\mu \nu}$ is the EMT. The Bianchi identity guarantees that energy is conserved: 
\bea
\label{e:conslaw}
\nabla_{\mu} G^{\mu \nu} =\nabla_{\mu} T^{\mu \nu }=0\,.
\eea
Mathematically, we can have all sorts of solutions to the above equations including those space-times which are not physically realistic. For example, they can have closed timelike curves or unstable negative energy density. That is where energy conditions come in, which provide constraints on allowable space-times.
Among the most commonly used are the Weak Energy Condition, Dominant Energy Condition, Strong Energy Condition, and the Null Energy Condition (NEC). The significance of the NEC is due to its role in the causal structure of space-time. It ensures light ray focusing, and underlies key theorems in GR, specifically singularity theorems. Also, other specified conditions are readily violated in the Concordance Model of Cosmology, while the NEC does not. It states that 
\bea
\label{e:nec}
T_{\mu \nu }k^\mu k^\nu \geq 0\,,
\eea
for any null vector $k^{\mu}$.
Singularity theorems are largely based on the Raychaudhuri equation, which for null geodesics reads, 
\bea
\label{e:raychau}
\frac{\mathrm{d}\theta }{\mathrm{d}\lambda} = -\frac{1}{d-2}\theta^2  - \sigma_{\mu \nu} \sigma^{\mu \nu} +\omega_{\mu \nu}\omega^{\mu \nu} - R_{\mu \nu}k^\mu k^\nu \,,
\eea
where $\lambda$ is the affine parameter parameterizing null geodesics, $\theta$ is the expansion scalar, $\sigma_{\mu \nu}$ is the shear tensor, $\omega_{\mu \nu}$ the vorticity tensor, and $R_{\mu \nu}$ the Ricci tensor.

In the absence of vorticity, assuming the Null Curvature Condition, 
\bea
\label{e:ricci}
R_{\mu \nu} k^\mu k^\nu \geq 0\,.
\eea
will result in a diverging expansion scalar $\theta$ at finite $\lambda$. 
Once again, we do not have a guide as to what constitutes a reasonable or unreasonable Ricci scalar.
Nevertheless, through Einstein’s equations $R_{\mu \nu} k^\mu k^\nu=8\pi G T_{\mu \nu}k^{\mu}k^{\nu}$. Hence, we can replace the Null Curvature Condition with the Null Energy Condition, which seems to be fulfilled for all classical known forms of energy.

Even before discussing singularities, the NEC ensures that light rays (in the absence of vorticity) can never de-focus. They either converge or stay parallel. This is the classical focusing theorem:
\bea
\label{e:theta1}
\frac{\mathrm{d}\theta}{\mathrm{d}\lambda}\leq 0\,, \quad \text{with} \quad \theta = \frac{1}{\mathcal{A}}\frac{\mathrm{d}\mathcal{A}}{\mathrm{d}\lambda}\,,
\eea
where $\mathcal{A}$ is an infinitesimal cross-sectional area element of the congruence. 
The locally (pointwise) energy conditions mentioned above, can be violated by quantum field effects \cite{Epstein:1965zza}. Also well-known is the Casimir effect, where quantum fluctuations produce negative energy densities between plates \cite{roman1986quantum}. This suggests that some other non-local forms of energy conditions may be more robust in these cases or circumstances. Instead of looking at energy at a single point, we can average it along an entire light path. This leads to the Average Null Energy Condition (ANEC). It states that for every inextendible null geodesic curve with affine parameter $\lambda$ and the tangent $k^\mu$,
\bea
\label{e:anec}
\int_{-\infty}^{\infty} \mathrm{d}\lambda T_{\mu\nu} k^\mu k^\nu \geq 0\,.
\eea
ANEC was proven and shown to hold in various cases \cite{yurtsever1995averaged,faulkner2016modular,yurtsever1990does,klinkhammer1991averaged,wald1991general,ford1996averaged,kontou2013averaged,hartman2017averaged}, and a milder version is the smeared energy condition \cite{freivogel2018smeared}. 
 
However, such conditions require a revision of singularity theorems based on the Raychaudhuri equation, given that the equation is a local one. 
The Quantum Null Energy Condition (QNEC) \cite{bousso2016quantum} is such a local condition. It incorporates quantum corrections systematically, connecting energy to entanglement and information. QNEC is a correction to the NEC which holds true in quantum field theories. Suppose we have a Cauchy surface $\Sigma$ and a co-dimension-two surface $\sigma$ that splits it into regions, one of which we call region $\mathcal{R}$. The second variation of the entropy of region $\mathcal{R}$, as the surface $\sigma$ is deformed in the null direction $k$, gives a local bound on the contracted EMT. This is called the QNEC, whose statement is as follows:
\bea
\label{e:qnec}
\langle T_{kk} \rangle \geq \frac{\hbar }{2 \pi \sqrt{h}}  S^{\prime\prime}_{\rm out}[\sigma]\,,
\eea
where prime denotes a derivative with respect to the affine parameter, $S_{\rm out}$ is the von Neumann entropy of region $\mathcal{R}$, $h$ is the determinant of the induced metric on the surface $\sigma$, and $ \langle T_{kk} \rangle= \langle T_{\mu \nu}k^{\mu}k^{\nu} \rangle$, with $\langle \rangle$ denoting the quantum expectation value for a given state. This condition was first derived from the quantum focusing conjecture (QFC) \cite{bousso2016quantum}. The QFC states that the "diagonal" quantum expansion of null geodesics is negative semi-definite $\Theta' \leq 0$. The  QNEC is then achieved by considering geodesics with vanishing classical null expansion, shear and vorticity $\theta=\sigma_{\mu \nu} \sigma^{\mu \nu}=\omega_{\mu \nu}\omega^{\mu \nu}=0$. 
The derivation of the QNEC is rather interesting. It started by considering a property of semi-classical gravity, the QFC, but after taking vanishing expansion, shear and vorticity, yielded a field theory statement that is independent of gravity---Newton's constant $G$ does not appear in it. Since then, the validity of the QNEC has been proven in  \cite{koeller2016holographic,bousso2016proof,malik2020proof,ceyhan2020recovering,Hollands:2025glm,balakrishnan2019general,panebianco2021loop}, and is independent of the validity or refutation of the QFC. The QFC simply served as a crucial guiding principle to propose the QNEC.

The limitation of vanishing null expansion is a rather severe one, since geodesics can focus or defocus also in flat spacetime given appropriate initial conditions.
In the holographic proof for CFTs, the QNEC was transformed using a conformal transformation, yielding conformal anomalies, and a condition with non-trivial scalar expansion, dubbed the holographic QNEC \cite{koeller2016holographic}:
\bea
\langle T_{\rm kk}   -A^{(T)}_{\rm kk} \rangle \geq \frac{\hbar}{2\pi \sqrt{h}}\left[ \left(S_{\rm out}-A^{(S)}\right)^{\prime \prime} + \frac{2}{d-2}\theta \left(S_{\rm out}-A^{(S)}\right)^{\prime}\right]\,,
\eea
where $A^{(T)},A^{(S)}$ are due to the anomalies coming from the conformal transformation of observables, and $S_{\rm out}$ should be interpreted as the finite part of the von Neumann/entanglement entropy.
The holographic QNEC is a partial step towards an energy condition with non-trivial expansion, but is not a different energy condition.
Considering 2D flat space-times, a more stringent energy condition preceded the $D>2$ QNEC: 
\be
\langle T_{kk}\rangle \geq \frac{\hbar}{2\pi} \left(S_{\rm out}^{\prime \prime}+\frac{6}{c}S_{\rm out}^{\prime 2}\right)\,,
\ee
where $c$ is the central charge of the CFT. Hence, it is tempting to consider other energy conditions that are stronger than the QNEC, which do not necessitate vanishing scalar expansion, and can include also $S_{\rm out}'$ terms.

We can once again think of the QFC, or better yet, the restricted QFC (rQFC), which says that at points where $\Theta=0$, $\Theta'\leq 0$ \cite{Shahbazi-Moghaddam:2022hbw}. The rQFC was proven in brane-worlds \cite{Shahbazi-Moghaddam:2022hbw}, and passed several tests \cite{Franken:2025gwr}. Another reason the rQFC is interesting is that a quantum extremal surface corresponds to a hypersurface with $\Theta=0$, rather than a single point. The results obtained using the QFC, and its implications for quantum extremal surfaces are still valid if only the rQFC is correct \cite{engelhardt2021world,hartman2020islands,ben2023islands,bousso2022islands,ben2025islands,akers2020quantum}.
Considering this limit of vanishing quantum expansion $\Theta=0$ results in an improved quantum null energy condition (INEC) which is again independent of $G$ \cite{ben2024quantum}: 
\begin{eqnarray}
\label{e:inec}
\langle T_{\rm kk} \rangle \geq \frac{\hbar}{2\pi \sqrt{h}}\left( S^{\prime \prime}_{\rm out} - \frac{d-3}{d-2}\,\rm \theta \, S^\prime_{\rm out} \right)\,.
\end{eqnarray}
The INEC has the desired properties of non-zero classical expansion, and the first derivative of the entropy $S_{\rm out}'$. Since the second term can be positive or negative, it is a stronger condition than the QNEC. 
In this paper, we analyze the INEC for conformal field theories (CFT) on a light cone/spherical entangling region. Since it will be based on a conformal transformation, anomaly terms will arise here too. 
 The analysis is based on the modular Hamiltonian technique \cite{kudler2025covariant}. We shall see that the INEC holds provided we require an additional constraint on the quantum state, similar to the smeared energy condition $2\pi  \,\int_0^{\gamma}d\lambda \, \lambda^d/\gamma^2\langle T_{\lambda \lambda}\rangle \geq  -\left\{\gamma\frac{ \delta^2 S_{\rm rel} }{\delta \gamma^2} -(d-3)\frac{\delta S_{\rm rel} }{\delta \gamma}\right\}/(d-1)$. 
 
We start by revisiting QFC and the proof of QNEC in section \ref{sec:qfc} to set the stage. We then present the formulation of the INEC in section \ref{sec:inec}. In section \ref{sec:LC} we perform the conformal transformation to the past light-cone,
and in \ref{sec:proof} we reduce the INEC to the smeared positivity condition above. In
\ref{ss:cylinder} we make the above condition a well-defined operator on the Lorentzian cylinder, verify
the INEC for the thermal state at leading order, and reduce the general case to a
monotonicity condition on the cap entropy. Finally, we suggest using the QNEC or INEC as the basis for semi-classical gravity which results in modified quantum focusing conjectures where the validity of the QNEC/INEC is guaranteed in \ref{sec:discussion}. We then conclude.

\section{QFC and proof of QNEC}
\label{sec:qfc}
To formulate the QFC \cite{bousso2016quantum}, let us first introduce the notion of quantum expansion which measures the response of generalized entropy under null deformations of a surface. 
Consider a codimension-two surface $\sigma$ of area $A$ that splits the Cauchy surface $\Sigma$ into two parts. We erect an orthogonal null hypersurface $N$ at $\sigma$. Let $\Lambda$ denote the affine parameter along the null generators. $\sigma$ is deformed along the null generators by an affine parameter amount $\Lambda= \Gamma(Y)$, where $Y$ are the transverse coordinates on $\sigma$. Any such deformed surface splits the Cauchy surface into two regions \cite{bousso2015new}. Let us define a one-parameter family of cuts given by
\begin{eqnarray}\label{e:gammaparam1}\Gamma(Y;\Lambda) &\equiv &\Gamma(Y;0) + \Lambda  \dot{\Gamma}(Y),\end{eqnarray}
with $\dot{\Gamma}(Y)>0$ and $\Gamma(Y;0)$ corresponding to the original surface $\sigma$. $\Lambda$ increases away from $\sigma$. The generalized entropy \cite{Bekenstein:1973ur,tHooft:1984kcu,Susskind:1994sm,dong2014holographic,Maldacena:1997re,engelhardt2015quantum,dong2018entropy} is defined as 
\begin{eqnarray}
\label{e:nec3}
S_{\rm gen}[\Gamma(Y)] &=&S_{\rm out}[\Gamma(Y)] +\frac{A[\Gamma(Y)]}{4G \hbar}\,.
\end{eqnarray}
Here, $S_{\rm out}$ is the von Neumann entropy of fields on one side of $\sigma$. The change in generalized entropy is given as
\bea
\label{e:changesgen}
\frac{\mathrm{d}S_{\rm gen}}{\mathrm{d}\Lambda}= \int_\Sigma \mathrm{d}^{d-2}Y \frac{\delta S_{\rm gen}}{\delta \Gamma(Y)}\dot{\Gamma}(Y) \equiv \frac{1}{4G\hbar}\int_\Sigma \mathrm{d}^{d-2}Y \sqrt{h}\Theta[\Gamma(Y),Y] \,.
\eea
Here, the quantum expansion at a point $Y_1$ is defined as the rate at which generalized entropy changes under a small variation of $\sigma$ along $N$, per cross-sectional area $\mathcal{A}$ of the variation:
\begin{eqnarray}
\label{e:nec5}
\Theta[\Gamma(Y);Y_1]= \frac{4G \hbar \,\dot{\Gamma}(Y)}{\sqrt{h}} \frac{\delta S_{\rm gen}[\Gamma(Y)]}{\delta \Gamma(Y_1)} \,,
\end{eqnarray}
where $h$ denotes the determinant of the intrinsic metric of $\sigma$. In the limit $\hbar \rightarrow 0$, this quantity reduces to classical expansion.  
The QFC states that quantum expansion cannot increase under a second variation of $\sigma$ along the same future null congruence:
\begin{eqnarray}
\label{e:nec6}
\frac{\mathrm{d}\Theta}{\mathrm{d}\Lambda}\leq 0\,.
\end{eqnarray}
In the limit where shear, vorticity, and classical expansion vanish, the above relation reduces to the QNEC:
\bea
\label{e:qnecdef}
\frac{2 \pi}{\hbar} \langle T_{\Lambda \Lambda} \rangle \geq  \frac{\delta^2 S_{\rm out}[\Gamma(Y)]}{\delta \Gamma(Y)^2}\,.
\eea
It is a lower bound on $\langle T_{k k} \rangle$ at a single point on the hypersurface $\sigma$. 
\begin{figure}
\centering
	\includegraphics[scale=0.6]{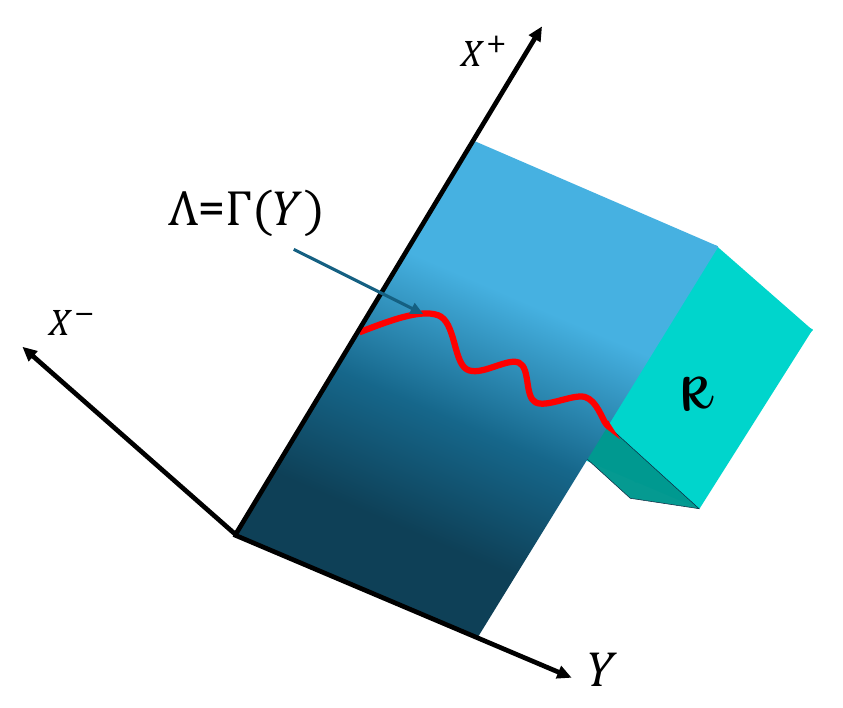}
	\caption{\em\small 
    A section of the null plane defined by $X^-=0$, with light-cone coordinates $X^{\pm}= X^1 \pm X^0$.  Here, $Y$ denotes the transverse coordinates $(X^2,X^3,\dots X^{d-1})$.  Let us consider a $d-2$ dimensional spatial surface $\Gamma$ on this null plane, given by the equation $\Lambda=\Gamma(Y)$ with $\Lambda$ being the affine parameter along $X^+$-- direction.  The modular Hamiltonian for region $\mathcal{R}$ bounded by this surface on the null plane is given in Eq.~\eqref{e:nec18}.
		\label{fig:v2}}
\end{figure}
The inequality has been proven in various cases with various techniques \cite{bousso2016proof,koeller2016holographic,balakrishnan2019general,malik2020proof,moosa2021renyi,roy2023proof, kudler2025covariant, Hollands:2025glm}. 

Let us briefly summarize the approach of \cite{kudler2025covariant} to proving the QNEC. Let $|\phi \rangle$ and $|\omega \rangle$ be two cyclic-separating vector states for the von Neumann algebra $\mathcal{B}$ and its causal complement $\mathcal{B^\prime}$ of quantum fields  in a causally complete
spacetime subregion. Assume the modular Hamiltonian decomposes as $H_\omega = H_\omega^\mathcal{B} + H_\omega^\mathcal{B^\prime}$, where $H_\omega^\mathcal{B}$ and $H_\omega^\mathcal{B^\prime}$ are the Hermitian forms affiliated with $\mathcal{B}$ and $\mathcal{B^\prime}$. If $|\phi \rangle$ is in the domain of $H_\omega^\mathcal{B}$ and the relative entropy is finite, then
\begin{eqnarray}
\label{e:nec13}
\Delta S_{\rm out}(\phi, \omega)=\langle \phi |H_\omega^\mathcal{B} |\phi \rangle - S_{\rm rel}(\phi || \omega)\,.
\end{eqnarray}
The modular Hamiltonian \cite{wall2012proof,casini2017modular} for the region $\mathcal{R}[\Gamma(Y)]$ above an
arbitrary cut $\Lambda=\Gamma(Y)$ of a null plane can be expressed as a local integral of the stress tensor (see fig. \ref{fig:v2}):
\begin{eqnarray}
\label{e:nec18}
H_\omega^\mathcal{B} = 2\pi \int_{\mathbb{R}^{d-2}} d^{d-2}Y \int_{\Gamma(Y)}^{\infty} d\Lambda (\Lambda-\Gamma(Y))T_{\Lambda \Lambda}(\Lambda,Y)\,.
\end{eqnarray}
Here, $\Lambda$ is an affine null coordinate, and $Y$ denotes the transverse coordinates.
Taking functional derivatives of the modular Hamiltonian, and using that the relative entropy is a convex function, i.e., the second shape derivative of the relative entropy is positive, one obtains the following result:
\begin{eqnarray}
\label{e:nec14}
2 \pi \langle \phi |T_{\Lambda \Lambda}(Y) |\phi \rangle \geq \frac{\delta^2 \Delta S_{\rm out}(\phi,\omega)}{\delta \Gamma(Y)^2}\,,
\end{eqnarray}
which reproduces the QNEC after setting $\hbar=1$, which we will do for the rest of the manuscript. 
%
\section{Improved Quantum Null Energy Condition}
\label{sec:inec}
Let us begin again with the quantum expansion
\bea
\label{e:t}
\Theta[\gamma(y)]= \frac{4G}{\sqrt{h}}\frac{\delta S_{\rm gen}[\gamma(y)]}{\delta \gamma(y)}\,.
\eea
Here, $\gamma(y)$ is the cut and $s$ is the parameter along the null direction.
After substituting the expression for the generalized entropy in the above equation, it takes the following form:
\bea
\label{e:inec1}
\Theta &= &\theta + \frac{4G}{\sqrt{h}}\frac{\delta S_{\rm out}}{\delta \gamma(y)}\bigg|_s\,,
\eea
where $\theta = \partial_s \log \sqrt{h}$ is the classical expansion. The rQFC is a statement about points where 
the quantum expansion vanishes, so we get the following relation:
\bea
\label{e:thetasout}
\Theta=0 \quad \Rightarrow \quad  \theta = - \frac{4G}{\sqrt{h}}\frac{\delta S_{\rm out}}{\delta \gamma(y)}\bigg|_s\,.
\eea
The (r)QFC then states, 
\bea
\label{e:qfc3}
\frac{d \Theta}{d s} = \frac{d \theta}{d s} - \theta \frac{4G }{\sqrt{h}} \frac{\delta S_{\rm out}}{\delta \gamma(y)}\bigg|_s + \frac{4G}{\sqrt{h}} \frac{d}{d s} \left(\frac{\delta S_{\rm out}}{\delta \gamma(y)}\bigg|_s \right) &\leq& 0\,.
\eea
Using Raychaudhuri’s equation and Einstein
field equations in the above for zero shear and vorticity, we get
\bea
\label{e:qfc4}
\frac{d \Theta}{d s} = -\frac{1}{d-2}\theta^2  - 8\pi G\langle T_{s s} \rangle  - \theta \frac{4G}{\sqrt{h}} \frac{\delta S_{\rm out}}{\delta \gamma(y)}\bigg|_s + \frac{4G}{\sqrt{h}} \frac{d}{d s} \left(\frac{\delta S_{\rm out}}{\delta \gamma(y)}\bigg|_s \right) &\leq & 0 \,,
\eea
where $\langle T_{s s} \rangle$ is the expectation value of the stress-energy tensor and $d$ is the space-time dimension. Substituting \Eqn{e:thetasout} in the above inequality, we get
\be
\label{e:qfc5}
 \frac{\theta}{d-2}\frac{4G}{\sqrt{h}}\frac{\delta S_{\rm out}}{\delta \gamma(y)}\bigg|_s  - 8\pi G\langle T_{s s} \rangle  - \theta\frac{4G}{\sqrt{h}} \frac{\delta S_{\rm out}}{\delta \gamma(y)}\bigg|_s+ \frac{4G}{\sqrt{h}} \frac{d}{d s} \left(\frac{\delta S_{\rm out}}{\delta \gamma(y)}\bigg|_s \right)\leq 0 \,,
 \ee%
 \be
\label{e:inec30}
\Rightarrow \boxed{ \frac{1}{\sqrt{h}}\frac{d}{d s} \left(\frac{\delta S_{\rm out}}{\delta \gamma(y)}\bigg|_s \right) - \frac{1}{\sqrt{h}}\left(\frac{d-3}{d-2}\right)\theta\frac{\delta S_{\rm out}}{\delta \gamma(y)}\bigg|_s  \leq 2\pi \langle T_{ ss}  \rangle \,.}
\ee
Here, the last inequality is referred to as the INEC \cite{ben2024quantum}. Similarly to the QNEC, $G$ does not appear in the energy condition anymore. Even though it originated from semiclassical gravity, it is a statement about field theories in Minkowski space-time.
We wish to stress that we have not made any approximations in the derivation such as neglecting higher order terms in $\theta$. Hence, regardless of how this inequality was conjured, the only question remaining is whether this is a correct statement in field theory. We shall show that the statement is actually correct, giving a genuine energy condition that extends the QNEC to non-vanishing classical expansion. A similar inequality for the derivatives of entanglement entropy of sphere of radius R was proven in \cite{casini2017theorem, Casini:2023kyj} using SSA, Lorentz invariance, and the Markov property of a CFT. It states that  $R \Delta S''_{out}(R)-(d-3) \Delta S'_{out}(R)<0$, where $\Delta S(R)$ corresponds to the difference in entanglement entropy between the CFT of two theories.
%
%
\section{Light Cone/ Sphere: Geometry and Modular Hamiltonian }
\label{sec:LC}
One technical point in QNEC is that the null surface has to have zero expansion at the point where the variation is made. In the proof, they achieved this by taking the null plane \cite{kudler2025covariant}. To prove the INEC, we need a null surface with non-zero expansion.
One such surface is the light cone. Consider light cones from a point in flat space:
Outgoing and incoming light rays from an event spread out or converge to an event with the area of spherical cross-sections growing like $A \sim r^2 $, with the scalar expansion $\theta = \pm(d-2)/r$, depending on whether the congruence is diverging or converging.

Let us first describe the setup \cite{rosso2020global,casini2018all,casini2017modular}. Given a $d$-dimensional Minkowski space-time $X^\mu$ with the metric signature $(-+\dots +)$. Consider a conformal transformation mapping the Minkowski space-time $X^\mu = (T,X,\vec{X}_\perp)$ into itself $x^\mu=(t,x,\vec{x}_\perp)$ given by the following relation:
\begin{eqnarray}
\label{e:cft}
    x^\mu = \frac{X^\mu + (X \cdot X)C^\mu }{ 1+ 2(X \cdot C)+ (X \cdot X)(C \cdot C)} - D^\mu\,,\quad \text{with} \quad D^\mu =\,(R,R,\vec{0})\,, \quad 
    C^\mu = \,(0,1/2R, \vec{0})\,.
\end{eqnarray}
Here, $C^\mu$ is the special conformal transformation parameter, $D^\mu$ is a space-time translation direction, and $X \cdot X =\,\, \eta_{\mu \nu} X^\mu X^\nu$. Define the null coordinates as 
\begin{eqnarray}
\label{e:lightcone}
    r^{\pm} = r \pm t\,, \quad \text{with} \quad t=x^0\,, \quad \text{and} \quad r= \sqrt{(x^1)^2 + \dots + (x^{d-1})^2}\,.
\end{eqnarray}
The map,  Eq.~\eqref{e:cft} takes the null plane and folds its infinitely extended null generators into null rays emanating from the cone's apex.  
The origin of the null-plane coordinates $X^\mu = 0$ is mapped to the point $x^\mu=(-R,-R,\vec{0})$ on the cone. Hence, the time coordinate $x^0=-R <0$ lies below the apex. The null plane $X^- = 0$ is mapped to the past light cone. The surface $X^\pm =0$, i.e., the intersection of both null planes, is mapped to the sphere $x^0=-R$, $r=R$. This is the spatial boundary of the cone. The transverse coordinates get mapped stereographically onto this sphere. And the points sent to large $X^+$ (with vanishing transverse displacement) collapse to the apex direction. The points on the null cone are labelled by $(\lambda,\Omega)$ after the conformal map.

\begin{figure}
\centering
	\includegraphics[width=0.6\linewidth, height =0.4\linewidth]{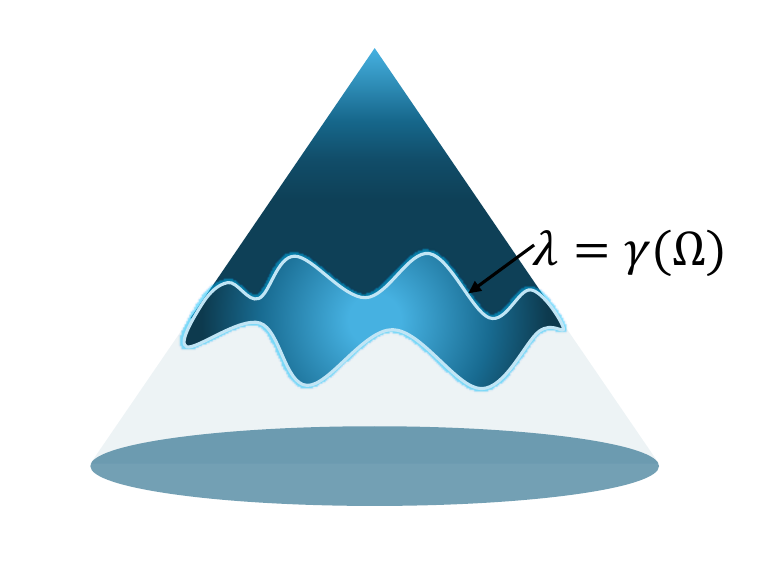}
	\caption{\em\small The past null cone, and the different regions. The vertical axis of the cone gives the time direction. The arbitrary region has boundary $\lambda = \gamma(\Omega)$ on the null cone. 
		\label{fig:v3}}
\end{figure}

Consider an entangling surface $\lambda= \gamma(\Omega)\geq 0 $ on the past light-cone, which represents the boundary of some region $\mathcal{R}(\lambda)$ with reduced density matrix $\rho_\mathcal{R}$ (see fig. \ref{fig:v3}). Here, $\lambda$ is the new affine parameter defined as
\bea
\label{e:newaffine}
\lambda= \frac{R p(\Omega)}{\Lambda + p(\Omega)}\,,\quad \text{where} \quad p(\Omega) = \frac{|\vec{x}_\perp(\Omega)|^2 + 4R^2}{4R}>0\,. 
\eea
and $\Omega$ denotes the transverse coordinates. The cut is a $(d - 2)$ sphere with radius $\gamma(\Omega)$:
\begin{eqnarray}
    ds^2  &=& \gamma(\Omega)^2 d\Omega^2_{d-2} \,.
\end{eqnarray}
Here, $d\Omega^2_{d-2} $ is the metric of a sphere $S^{d-2}$ of unit radius. The conformal transformation between the cut on the cone and plane is given by the following relation:
\bea
\label{e:cut}
\gamma &=& \frac{p(\Omega)R}{\Gamma + p(\Omega) }\,.
\eea
Now, consider the transformation of the EMT. The projected EMT along the null path on a null plane is 
\begin{eqnarray}
    \label{e:tmu}
    T_{\Lambda \Lambda} &=& \frac{dX^\mu}{d\Lambda}\frac{dX^\nu}{d\Lambda}T_{\mu \nu}\,,
\end{eqnarray}
with $\Lambda$ being the affine parameter on the plane. The transformation of this stress tensor when the Hilbert space $\mathcal{H}$ associated with the field theory in Minkowski is mapped to the Hilbert space $\mathcal{\bar{H}}$ of the transformed CFT using the unitary operator $U: \mathcal{H} \rightarrow \mathcal{\bar{H}}$ is
\begin{eqnarray}
    \label{e:teant}
    U T_{\Lambda\Lambda}(\Lambda, \vec{Y})U^\dagger &=& |\omega(\Lambda,Y)|^{2-d} (\bar{T}_{\Lambda \Lambda}(\Lambda, \vec{Y})- \bar{S}_{\Lambda \Lambda})\,. 
\end{eqnarray}
 The anomalous term, $\bar{S}_{\Lambda \Lambda}$, ensures that the expectation value of mapped stress tensor vanishes when evaluated in the transformed vacuum state \cite{brown1977stress, herzog2013stress}. It is non-vanishing for even $d$.
To reduce clutter, we remove the bar notation, and whenever written below it should be understood as the anomaly subtracted quantities. 
We can now write the modular Hamiltonian for the past null cone with a cut at $\lambda =\gamma(\Omega)$ \cite{casini2017modular,rosso2020global}:
\begin{eqnarray}
\label{e:cone_ham}
    H &=& 2\pi \int d\Omega_{d-2} \int_0^{\gamma(\Omega)} d\lambda  \, \lambda^{d-1} \frac{\gamma(\Omega)- \lambda}{\gamma(\Omega)} T_{\lambda \lambda} \,,
\end{eqnarray}
with $\lambda$ being an affine coordinate over the null rays defined in Eq.~\eqref{e:newaffine}. 
\section{Towards proving the INEC}
\label{sec:proof}
Consider a past light cone in $d$-dimensional Minkowski space-time. Take the cut $\lambda=\gamma(\Omega)$, $d-2$ dimensional spatial cross-section of the light cone, to be the entangling surface. $\Omega$ denotes the transverse coordinates on the cone. There are four null hypersurfaces orthogonal to the cut $\gamma(\Omega)$, two of them (future/past directed ingoing) have negative classical expansion and the other two (future/past directed outgoing) have positive expansion. The modular Hamiltonian that generates the boost along the null direction for this entangling surface is given in \Eqn{e:cone_ham}. Consider one-parameter family of cuts
\begin{eqnarray}
\label{e:gammaparam}
    \gamma (\Omega;s) &\equiv& \frac{\gamma(\Omega;0)}{1+ s \,\alpha(\Omega)}\,.
\end{eqnarray}
Here, $s$ is along the future-directed null geodesic and $\alpha(\Omega)>0$ is the acceleration parameter. $s=0$ on the cut $\gamma(\Omega;0)$ and increases as we move along the null hypersurface. The classical expansion is given as
\bea
\label{e:clasex}
\theta = \partial_s \log \sqrt{h} = \frac{d-2}{\gamma(\Omega)}\dot{\gamma}(\Omega) \,,
\eea
where $h_{a b}= \gamma^2 g_{ab}$ is the induced metric on the surface and $g_{ab}$ is the metric of unit $(d-2)$ sphere. 

Let $\rho_\mathcal{R}$ be the reduced density matrix of the enclosed region. Let $\sigma_\mathcal{R}$ be the vacuum state, the relative entropy is then defined as
\bea
\label{e:relentdef}
S_{\rm rel}(\rho_{\mathcal{R}}|| \sigma_{\mathcal{R}})= \rm{Tr}(\rho \ln \rho )- \rm{Tr}(\rho \ln \sigma)\,.
\eea
Then the monotonicity of the relative entropy $S_{\rm rel}(\rho_{\mathcal{R}}|| \sigma_{\mathcal{R}})$ implies
\bea\label{e:firstderrel1}
\frac{\delta S_{\rm rel}}{\delta \gamma}\geq 0\,.
\eea
Convexity of relative entropy as proven in \cite{ceyhan2020recovering} states that if we vary the region boundary along the null generators, the second derivative of relative entropy must be positive:
\bea
\label{e:secderrel1}
\frac{\mathrm{\delta}^2 S_{\rm rel}(\rho_{\mathcal{R}}|| \sigma_{\mathcal{R}})}{\mathrm{\delta} \Gamma^2}\geq 0 \,,
\eea
for the cut $\Gamma$ on plane. Using the conformal relation \Eqn{e:cut}, we get
\bea
\label{e:srerel}
\frac{\mathrm{\delta}^2 S_{\rm rel}}{\mathrm{\delta} \gamma^2} + \frac{2}{\gamma}\frac{\delta S_{\rm rel}}{\delta \gamma}\geq 0 \,.
\eea
Let's start by writing the INEC again
\bea\label{e:inecstate}
\Delta S^{\prime \prime} - \left(\frac{d-3}{d-2}\right)\theta \Delta S^\prime \leq 2\pi \sqrt{h}\langle T_{\lambda \lambda}  \rangle \,,
\eea
and consider the vacuum-subtracted modular Hamiltonian associated with this region,
\begin{eqnarray}
\label{e:cone_ham1}
   \Delta \langle H \rangle &=& 2\pi \int d\Omega^\prime_{d-2} \int_0^{\gamma(\Omega^\prime)} d\lambda  \, \lambda^{d-1} \frac{\gamma(\Omega^\prime)- \lambda}{\gamma(\Omega^\prime)} \langle T_{\lambda \lambda} \rangle\,.
\end{eqnarray}
Taking the first functional derivative of the above equation, we get
\bea
\label{e:firfunder}
\frac{\delta \Delta \langle H \rangle }{\delta \gamma(\Omega)}=2\pi \int d\Omega^\prime_{d-2} \delta(\Omega^\prime-\Omega)\int_0^{\gamma(\Omega^\prime)}d\lambda \frac{\lambda^d}{\gamma(\Omega^\prime)^2}\langle T_{\lambda \lambda}(\lambda,\Omega^\prime)\rangle 
=2\pi \,\int_0^{\gamma(\Omega)}d\lambda \frac{\lambda^d}{\gamma(\Omega)^2}\langle T_{\lambda \lambda}(\lambda,\Omega)\rangle\,.
\eea
And the second functional derivative is
\bea
\label{e:secfunder}
\frac{\delta^2 \Delta \langle H \rangle }{\delta \gamma(\Omega)^2} = 2\pi \, \gamma(\Omega)^{d-2} \langle T_{\lambda \lambda}[\gamma(\Omega),\Omega] \rangle - 2\pi  \int_0^{\gamma(\Omega)} \mathrm{d}\lambda \, \frac{2 \lambda^d}{\gamma(\Omega)^3}\, \langle T_{\lambda \lambda}(\lambda,\Omega) \rangle\,.
\eea
We have the following relation between the von Neumann entropy, the relative entropy, and vacuum-subtracted modular
Hamiltonian:
\bea
\label{e:relmod}
 \Delta S_{\rm out} = \Delta \langle H \rangle - S_{\rm rel}\,.
\eea
Substituting \eqref{e:firfunder},\eqref{e:secfunder},\eqref{e:relmod}, in the LHS of the INEC, we get
\bea
\label{e:inecp1}
\frac{\delta^2 \Delta S_{\rm out} }{\delta \gamma(\Omega)^2} - \frac{d-3}{\gamma(\Omega)} \frac{\delta \Delta S_{\rm out} }{\delta \gamma(\Omega)} &=&\frac{\delta^2 \Delta \langle H \rangle }{\delta \gamma(\Omega)^2} - \frac{\delta^2 S_{\rm rel} }{\delta \gamma(\Omega)^2} - \frac{d-3}{\gamma(\Omega)}\left\{ \frac{\delta \Delta \langle H \rangle }{\delta \gamma(\Omega)} - \frac{\delta S_{\rm rel} }{\delta \gamma(\Omega)}  \right\}\,,
\\&=& 2\pi \, \gamma(\Omega)^{d-2} \langle T_{\lambda \lambda}[\gamma(\Omega),\Omega] \rangle - \nonumber 4\pi  \int_0^{\gamma(\Omega)} \mathrm{d}\lambda \, \frac{ \lambda^d}{\gamma(\Omega)^3}\, \langle T_{\lambda \lambda}(\lambda,\Omega) \rangle-  \nonumber \\&& 2\pi (d-3) \,\int_0^{\gamma(\Omega)}d\lambda \, \frac{\lambda^d}{\gamma(\Omega)^3}\langle T_{\lambda \lambda}(\lambda,\Omega)\rangle -\frac{\delta^2 S_{\rm rel} }{\delta \gamma(\Omega)^2} +\frac{d-3}{\gamma(\Omega)}\frac{\delta S_{\rm rel} }{\delta \gamma(\Omega)}\,.
\\
\label{e:inecp2}
\frac{\delta^2 \Delta S_{\rm out} }{\delta \gamma(\Omega)^2} - \frac{d-3}{\gamma(\Omega)} \frac{\delta \Delta S_{\rm out} }{\delta \gamma(\Omega)} &=& 
2\pi \, \gamma(\Omega)^{d-2} \langle T_{\lambda \lambda}[\gamma(\Omega),\Omega] \rangle - 2\pi (d-1) \,\int_0^{\gamma(\Omega)}d\lambda \, \frac{\lambda^d}{\gamma(\Omega)^3}\langle T_{\lambda \lambda}(\lambda,\Omega)\rangle \nonumber \\&-& \left\{\frac{\delta^2 S_{\rm rel} }{\delta \gamma(\Omega)^2} +\frac{2}{\gamma(\Omega)}\frac{\delta S_{\rm rel} }{\delta \gamma(\Omega)}\right\}+\frac{2}{\gamma(\Omega)}\frac{\delta S_{\rm rel} }{\delta \gamma(\Omega)}+\frac{d-3}{\gamma(\Omega)}\frac{\delta S_{\rm rel} }{\delta \gamma(\Omega)}\,.
\eea
 INEC will be true if the following is true
\bea
\label{e:con1}
\boxed{\frac{d-1}{\gamma(\Omega)}\left(2\pi \,\int_0^{\gamma(\Omega)}d\lambda \, \frac{\lambda^d}{\gamma(\Omega)^2}\langle T_{\lambda \lambda}(\lambda,\Omega)\rangle - \frac{\delta S_{\rm rel} }{\delta \gamma(\Omega)}\right)\geq - \left\{\frac{\delta^2 S_{\rm rel} }{\delta \gamma(\Omega)^2} +\frac{2}{\gamma(\Omega)}\frac{\delta S_{\rm rel} }{\delta \gamma(\Omega)}\right\} }\label{eq:smeared-cone}
\eea
where we deliberately separated the expressions on both sides of the above equation. The term in the curly bracket on the RHS is positive from \Eqn{e:srerel}. This means that the term on the LHS is bounded from below by a negative semi-definite term. Additionally, we can recognize the LHS as the first derivative of the von Neumann entropy, then the above condition can be phrased as:
\bea
 \label{e:con2} \boxed{\frac{d-1}{\gamma(\Omega)}\frac{\delta \Delta S_{\rm out}}{\delta \gamma(\Omega)}\geq  - \left\{\frac{\delta^2 S_{\rm rel} }{\delta \gamma(\Omega)^2} +\frac{2}{\gamma(\Omega)}\frac{\delta S_{\rm rel} }{\delta \gamma(\Omega)}\right\}},
 \eea
meaning that the first derivative of the von Neumann entropy is bounded from below by a negative semi-definite term.

\theoremstyle{plain}
\newtheorem{proposition}{Proposition}
\newtheorem{corollary}{Corollary}
\theoremstyle{remark}
\newtheorem{remark}{Remark}
To assess the physical content of the above integrated inequality \eqref{e:con1}, we develop the leading-order (first-law) analysis here in which the right-hand side drops out entirely. We consider a one-parameter family of states $\rho_\epsilon$ that are small perturbation of the vacuum $\sigma$:
\bea
\label{e:rhopert}
\rho_\epsilon = \sigma + \epsilon \delta\rho + O(\epsilon^2)\,,
\eea
where $\epsilon$ is a dimensionless small parameter and $\delta\rho$ denotes the first-order state variation, with $\mathrm{Tr}\delta \rho=0$. Relative entropy is non-negative and vanishes only at $\rho_{\epsilon=0} = \sigma$; along the smooth family $\rho_\epsilon$ it therefore attains its minimum at $\epsilon = 0$ with vanishing first derivative, so it is second order \cite{ceyhan2020recovering,blanco2013relative}.
\bea
\label{e:Srel-second-order}
S_{\rm rel}(\rho_\epsilon\|\sigma)=O(\epsilon^2),
\qquad\Rightarrow\qquad
\frac{\delta S_{\rm rel}}{\delta\gamma}=O(\epsilon^2)\,,
\quad
\frac{\delta^2 S_{\rm rel}}{\delta\gamma^2}=O(\epsilon^2)\,.
\eea
Combined with the exact relation \eqref{e:relmod}, the von Neumann entropy $S(\rho)$ and the modular Hamiltonian $H$ satisfy the first law of entanglement entropy:
\bea
\label{e:firstlaw}
\Delta S_{\rm out} = \Delta \langle H \rangle + O(\epsilon^2)\,.
\eea
\begin{proposition}[INEC in the first-law regime]
\label{prop:firstlaw}
For the family $\rho_\epsilon$ above, assuming that the leading-order null energy satisfies
\bea
\label{e:smeared-pos}
\int_0^{\gamma(\Omega)} d\lambda\,
\frac{\lambda^{d}}{\gamma(\Omega)^2}\,t_{\lambda\lambda}(\lambda,\Omega)\;\ge\;0\,,
\eea
where $\langle T_{\lambda\lambda}\rangle_{\rho_\epsilon} =
\epsilon\,t_{\lambda\lambda} + \mathcal{O}(\epsilon^{2})$, then the INEC holds to leading order in $\epsilon$.
\end{proposition}

\begin{proof}
By \eqref{e:Srel-second-order} the right-hand side of \eqref{e:con2} is $O(\epsilon^2)$,
\bea
\label{e:pro1}
  \frac{d-1}{\gamma}\,\frac{\delta\Delta S_{\rm out}}{\delta\gamma}
  \;\geq\; \mathcal{O}(\epsilon^{2}).
\eea
The left-hand side equals 
\bea
  \frac{d-1}{\gamma}\cdot 2\pi\epsilon\int_{0}^{\gamma}\!\frac{\lambda^{d}}{\gamma^{2}}\,
  t_{\lambda\lambda}\,d\lambda + \mathcal{O}(\epsilon^{2})\,,
\eea
which is non-negative at $O(\epsilon)$ by
\eqref{e:smeared-pos} (the prefactor $(d-1)/\gamma$ is positive). The inequality
\eqref{e:con1} therefore holds, and with it the INEC.
\end{proof}

\begin{remark}
\label{rem:weight}
The weight $w(\lambda)=\lambda^{d}/\gamma^2$ is strictly positive on $(0,\gamma)$ and
vanishes at the apex $\lambda=0$. Condition \eqref{e:smeared-pos} is thus a weighted
half-line null energy condition. It is weaker than pointwise NEC: pointwise positivity would certainly imply it, but is far from necessary. It is satisfied
whenever the positive null energy near the cut dominates any negative contribution
near the apex, after weighting by $\lambda^{d}$.
\end{remark}
Proposition~\ref{prop:firstlaw} is sharp: at leading order the INEC is
equivalent to \eqref{e:smeared-pos}. For \eqref{e:smeared-pos} to be a candidate theorem, the smeared integral must first exist as a well-defined operator and have finite vacuum fluctuations. The weight $w(\lambda)=\lambda^d/\gamma^2$ of the integral is zero at the apex $(\lambda=0)$ and $w(\lambda)=\gamma^{d-2}\neq 0$ at the cut $(\lambda=\gamma)$. Its left-hand side is the null energy
averaged over a single finite segment $[0,\gamma]$ against a weight that does not
vanish at the cut $\lambda=\gamma$.
Along a null ray, the stress-tensor correlator is dominated by intense short-distance fluctuations as two points converge, scaling as $(\lambda-\lambda')^{-2d}$ by the conformal dimension of the stress tensor ($\Delta_T=d$). If the weight applied to this measurement does not smoothly taper off at the boundaries, these vacuum fluctuations of the averaged operator diverge for $d>2$. As a result, the sharply-cut single-interval null average is not a well-defined operator: it is unbounded below \cite{freivogel2018smeared}. No positivity statement about it can be established
intrinsically. To obtain a well-posed setting in which \eqref{e:smeared-pos} becomes a
theorem, we pass to the Lorentzian cylinder.

\section{The INEC on the Lorentzian cylinder \texorpdfstring{$\mathbb{R}\times S^{d-1}$}{R x S(d-1)}}
\label{ss:cylinder}
On the light cone the smeared energy piece of \eqref{e:con1} is built from the null
energy averaged over a single null segment, an operator that is unbounded
below and has infinite vacuum fluctuations for $d>2$. We evade this by transitioning to the Lorentzian cylinder
$\mathbb{R}\times S^{d-1}$. In this geometry, the smeared integral is a well-defined, bounded operator, with a built-in weight that vanishes at the
endpoints of the (finite) geodesic. We use the vacuum modular Hamiltonian of a cap
region computed by \cite{rosso2020global} to write the INEC, and reduce it to a clean monotonicity condition.

On $\mathbb{R}\times S^{d-1}$, a null geodesic runs between antipodal points of the spatial
$S^{d-1}$ parametrized by (affine) $\beta\in[0,\pi]$, with transverse coordinates $\Omega \in S^{d-2}$. The light-cone cut at $\gamma$ is replaced by an entangling cut at $\beta=\bar A$ which defines a cap region \cite{rosso2020global}. The induced metric on the cut
is $R^2\sin^2\!\beta\,d\Omega_{d-2}^2$. Introducing a flow parameter $s$, such that $\bar A=\bar A(s)$, the
classical expansion of the congruence at the cut is 
\bea
 \label{eq:cyl-theta-gen}
  \theta \;=\; \partial_s\log\sqrt h\big|_{\rm cut}
  \;=\; \frac{d}{ds}\log\!\big[(R\sin\bar A)^{d-2}\big]
  \;=\; (d-2)\cot\bar A\;\dot{\bar A},
  \qquad \dot{\bar A}\equiv\frac{d\bar A}{ds}\,,
\eea
in exact parallel with the cone, where $\theta=\tfrac{d-2}{\gamma}\dot\gamma$, under
the dictionary $1/\gamma\leftrightarrow\cot\bar A$ and $\dot\gamma\leftrightarrow
\dot{\bar A}$. Since $\beta$ is affine, we chose 
the flow parameter $s$ to coincide with the affine parameter at the
cut, so that $\dot{\bar A}=1$, $\ddot{\bar A}=0$, then the classical expansion is 
\bea
\label{e:cyl-theta}
\theta=\partial_s\log\sqrt h=(d-2)\cot\bar A\,,
\eea
which is positive for $\bar A<\pi/2$, vanishes at the hemisphere $\bar A=\pi/2$, and reduces to the
light-cone value $(d-2)/\gamma$ as $\bar A\to0$. The relevant averaged null energy on the cylinder is the full geodesic operator
\bea
\label{e:cyl-operator}
\mathcal{N}=\int_0^{\pi}d\beta\,\sin^{d}\!\beta\,\Tbb(\beta)\geq0\,,
\eea
whose weight $\sin^{d}\!\beta$ vanishes smoothly (to order $d$) at both endpoints $\beta=0,\pi$ \cite{rosso2020global}. 

The one-sided cap modular Hamiltonian is
\cite{rosso2020global}
\bea
\label{e:cyl-K}
\Delta\langle K\rangle
=2\pi\int d\Omega_{d-2}\int_0^{\bar A(\Omega)}d\beta\,(R\sin\beta)^{d-2}\,
\frac{\sin\beta\,\sin(\bar A(\Omega)-\beta)}{\sin\bar A(\Omega)}\,\langle\Tbb(\beta,\Omega)\rangle\,,
\eea
where $\Tbb$ is the vacuum-subtracted null stress component. Working at a fixed transverse point $\Omega$, define the partial moments
\bea
\label{e:partmoments}
  m_k(\bar A) \;\equiv\;
  \int_0^{\bar A(\Omega)}\!d\beta\,(R\sin\beta)^{d-2}\sin^k\!\beta\;\langle\Tbb(\beta,\Omega)\rangle\,.
\eea
Differentiating \eqref{e:cyl-K}, and using
$\partial_{\bar A}\!\left[\sin(\bar A-\beta)/\sin\bar A\right]=\sin\beta/\sin^2\bar A$
together with the vanishing of the boundary term at $\beta=\bar A$,
\bea
  \label{eq:cyl-Kp2}
  \big[\Delta\langle K\rangle\big]'
  \;\equiv\;\frac{\delta\,\Delta\langle K\rangle}{\delta\bar A(\Omega)}
  &=& 2\pi\int d\Omega'_{d-2}\,\delta(\Omega'-\Omega)
      \int_0^{\bar A(\Omega')}\!d\beta\,(R\sin\beta)^{d-2}\,
      \frac{\sin^2\!\beta}{\sin^2\bar A(\Omega')}\,\langle\Tbb(\beta,\Omega')\rangle
  \nonumber\\
  &=& \frac{2\pi}{\sin^2\bar A}\,m_2(\bar A)\,.
\eea
A second derivative produces a local stress-tensor term at the cut,
\bea
\label{eq:cyl-Kpp2}
  \big[\Delta\langle K\rangle\big]''
  = -\frac{4\pi\cos\bar A}{\sin^3\bar A}\,m_2(\bar A)
    \;+\; 2\pi\,(R\sin\bar A)^{d-2}\,\,\langle\Tbb(\bar A)\rangle\,,
\eea
the second term being $2\pi\sqrt h\,\langle\Tbb(\bar A)\rangle$.
The convexity of relative entropy on the null plane \eqref{e:secderrel1}, carried to the
cylinder cut through the conformal relation between $\bar A$ and the flat cut $\Gamma$, takes the form
\bea
\label{e:cyl-convexity}
\frac{\delta^2 S_{\rm rel}}{\delta\bar A^2} + 2\cot\bar A\,\frac{\delta S_{\rm rel}}{\delta\bar A}
\;\geq\;0\,,
\eea
the cylinder analogue of \eqref{e:srerel}. Writing the INEC \eqref{e:inecstate} on the
cylinder, where $\theta=(d-2)\cot\bar A$ and $\sqrt h=(R\sin\bar A)^{d-2}$,
\bea
\label{e:cyl-inecstate}
[\Delta S_{\rm out}]'' - (d-3)\cot\bar A\,[\Delta S_{\rm out}]'
\;\leq\;2\pi(R\sin\bar A)^{d-2}\langle\Tbb(\bar A)\rangle\,,
\qquad{}'\equiv\frac{\delta}{\delta\bar A}\,,
\eea
and using $\Delta S_{\rm out}=\Delta\langle K\rangle-S_{\rm rel}$ with the shape derivatives
\eqref{eq:cyl-Kp2}, \eqref{eq:cyl-Kpp2}, the local term $2\pi\sqrt h\,\langle\Tbb(\bar A)\rangle$
cancels on both sides. The INEC then holds provided
\bea
\label{e:cyl-con1}
\boxed{(d-1)\cot\bar A\left(\frac{2\pi}{\sin^2\bar A}\,m_2(\bar A)
- \frac{\delta S_{\rm rel}}{\delta\bar A}\right)
\;\geq\;
-\left\{\frac{\delta^2 S_{\rm rel}}{\delta\bar A^2}
+ 2\cot\bar A\,\frac{\delta S_{\rm rel}}{\delta\bar A}\right\}}
\eea
the cylinder analogue of \eqref{e:con1}. The bracketed combination on the left is
$[\Delta\langle K\rangle]'-S_{\rm rel}'=[\Delta S_{\rm out}]'$, so equivalently
\bea
\label{e:cyl-con2}
\boxed{(d-1)\cot\bar A\,\frac{\delta\Delta S_{\rm out}}{\delta\bar A}
\;\geq\;
-\left\{\frac{\delta^2 S_{\rm rel}}{\delta\bar A^2}
+ 2\cot\bar A\,\frac{\delta S_{\rm rel}}{\delta\bar A}\right\}}
\eea
the analogue of \eqref{e:con2}. By \eqref{e:cyl-convexity} the right-hand side is negative
semi-definite, so for a sub-hemisphere cap ($\cot\bar A>0$) the INEC is controlled by the
first derivative of the vacuum-subtracted cap entropy, bounded below by a negative quantity.
\subsection{High temperature}
\label{ss:hight}
Here, we do a non-perturbative check of the general monotonicity condition
\eqref{e:cyl-con2}, valid in the high temperature regime where the state is far from the
vacuum. 
At high temperature the cap entanglement entropy is dominated by the thermal
(extensive) contribution. Let $s_{\rm th}>0$ be the thermal entropy density. Volume of the capped region is 
\bea
  V_{\rm cap}(\bar A)
  = \mathrm{Vol}(S^{d-2})\,R^{d-1}\!\int_0^{\bar A}\!\sin^{d-2}\!\beta\,d\beta\,.
\eea
The vacuum-subtracted entropy can be expressed in terms of thermal entropy, 
$\Delta S_{\rm out}\simeq s_{\rm th}\,V_{\rm cap}(\bar A)$, at high $T$, \cite{ben2023islands,ben2025islands}. Differentiating,
\bea
  [\Delta S_{\rm out}]'
  \;\simeq\; s_{\rm th}\,V_{\rm cap}'(\bar A)
  \;=\; s_{\rm th}\,\mathrm{Vol}(S^{d-2})\,R^{d-1}\,\sin^{d-2}\!\bar A
  \;>\;0\,,
  \label{a:dS-highT}
\eea
strictly positive for every $\bar A\in(0,\pi)$. Here, the monotonicity is a direct consequence of the extensivity
of thermal entropy and the monotonic growth of the cap volume. 
\subsection{Perturbative proof}
\label{sec:pertproof}
Eq~\eqref{e:cyl-con1} at leading order becomes the cylinder analogue of
\eqref{e:smeared-pos}, 
\bea
\label{e:cyl-smeared}
J_{\rm cyl}=\frac{2\pi(d-1)\cos\bar A}{\sin^3\bar A}\,m_2(\bar A)\ge0
\qquad\Longleftrightarrow\qquad
m_2(\bar A)\ge0
\quad(\bar A<\pi/2)\,.
\eea
Here, $J_{\rm cyl}$ is the smeared integral. Unlike its light-cone counterpart, the left-hand side here is a partial moment of the
well-defined operator \eqref{e:cyl-operator}; the full
moment $m_2(\pi)\!\propto\!\int_0^\pi\!\sin^d\!\beta\,\langle\Tbb\rangle\ge0$ is Rosso's 
cylinder bound \cite{rosso2020global}, with the weight $\sin^d\beta$ vanishing smoothly at both endpoints. 

The restriction to region, $A<\pi/2$, is geometric and without loss of generality. When we switch to region where our cut is at $A>\pi/2$, we consider the modular Hamiltonian for the complementary region,
\bea
\label{e:cyl-Kcomp}
\Delta\langle K_S\rangle
=2\pi\int d\Omega_{d-2}\int_{\bar A(\Omega)}^\pi d\beta\,(R\sin\beta)^{d-2}\,
\frac{\sin\beta\,\sin(\beta-\bar A(\Omega))}{\sin\bar A(\Omega)}\,\langle\Tbb(\beta,\Omega)\rangle\,,
\eea
whose shape derivative is $[\Delta\langle K_S\rangle]'=-\,2\pi\,n_2(\bar A)/\sin^2\bar A$ with the
complementary second moment
\bea
\label{e:cyl-nmoment}
n_2(\bar A)\;\equiv\;\int_{\bar A(\Omega)}^{\pi}\!d\beta\,(R\sin\beta)^{d-2}\sin^2\!\beta\,\langle\Tbb(\beta,\Omega)\rangle
\;=\;R^{d-2}\!\int_{\bar A}^{\pi}\!\langle\Tbb(\beta)\rangle\sin^d\!\beta\,d\beta\,.
\eea
For this cut, INEC at leading order takes the following form
\bea
\label{e:cyl-smeared-comp}
 J^{S}_{\rm cyl}
 = -\,\frac{2\pi(d-1)\cos \bar A}{\sin^3 \bar A}\,n_2(\bar A)\;\ge\;0
 \qquad\Longleftrightarrow\qquad
 n_2(\bar A)\;\ge\;0\,, \qquad (\bar A>\pi/2)\,,
\eea
the equivalence holding because for a super-hemisphere cap $\cos\bar A<0$, so the prefactor
$-\cos\bar A/\sin^3\bar A>0$.
The sign flip arises because $\bar A$ is now the lower endpoint.

\begin{corollary}[INEC for the thermal state on the cylinder, first-law regime]
\label{cor:thermal}
Let the CFT be in its thermal (Gibbs) state on $\mathbb{R}\times S^{d-1}$,
perturbatively above the vacuum. Then \eqref{e:cyl-smeared} holds for every
sub-hemisphere cap, and the INEC holds to leading order with no further assumption.
\end{corollary}

\begin{proof}
Consider a CFT on the spatial sphere $S^{d-1}$ of radius $R$ (the constant-time
slice of the cylinder), in the Gibbs state
\begin{equation}
  \rho_{\rm th} = \frac{e^{-\beta_{\rm th}H}}{Z},
  \qquad \beta_{\rm th} = 1/T,
\end{equation}
with $H$ the generator of cylinder time $\sigma$. Because $\rho_{\rm th}$ is static and $SO(d)$-invariant, its stress-tensor one-point
function is fixed by symmetry to the perfect-fluid form
\begin{equation}
  \langle T_{\mu\nu}\rangle_{\rm th}
  = (\rho_E+p)\,u_\mu u_\nu + p\,g_{\mu\nu},
  \qquad u^\mu=\partial_\sigma\,,
\end{equation}
with constant energy density $\rho_E$ and pressure $p$. Projecting onto the cylinder null geodesic (tangent $dx^\mu/d\beta=(-R,1,\vec0)$ in
$(\sigma,\theta,\vec x_\perp)$ coordinates), the vacuum-subtracted stress tensor is
\bea
  \label{eq:th-Tbb}
  \langle\Tbb(\beta)\rangle_{\rm th}
  = R^2(\bar\rho_E+\bar p)\;\equiv\;\mathcal{E}_0\;>\;0\,.
\eea
After substituting the above equation in \eqref{e:partmoments}, the partial moment reduces to 
\begin{equation}
  m_2(\bar A)
  = \int_0^{\bar A}\!d\beta\,(R\sin\beta)^{d-2}\sin^2\!\beta\,\langle\Tbb\rangle
  = R^{d-2}\mathcal{E}_0\,I_d(\bar A),
  \qquad
  I_d(\bar A)=\int_0^{\bar A}\!\sin^{d}\!\beta\,d\beta>0,
  \label{eq:th-m2}
\end{equation}
which is strictly positive because $\sin^{d}\!\beta>0$ on $(0,\bar A)\subset(0,\pi)$. $J_{\rm cyl}\sim\bar A^{\,d-2}$ matches the light-cone form in the
limit $\bar A\to0$, since the cylinder integrand $(R\sin\beta)^{d-2}\sin^2\beta \to R^{d-2}\beta^d$
and $\cos \bar A/\sin^3 \bar A \to 1/\gamma^3$ as $\bar A\to0$, reproducing the cone $J\sim\gamma^{d-2}$. Eq.~\eqref{e:cyl-smeared} then holds for any
sub-hemisphere cap ($\bar A<\pi/2$),
\bea
  J_{\rm cyl}=\frac{2\pi(d-1)\cos\bar A}{\sin^3\bar A}\,R^{d-2}\mathcal{E}_0\,I_d(\bar A)
  \;>\;0\,.
  \label{eq:th-J}
\eea
\end{proof}
Both of the above checks are in a regime where the vacuum-subtracted energy along the null direction is pointwise positive.
By Remark~\ref{rem:weight}, pointwise positivity already implies the weighted
condition \eqref{e:smeared-pos} (equivalently \eqref{e:cyl-smeared}). These are
therefore consistency checks rather than strong tests of the INEC. The weighted condition
becomes nontrivial only when the vacuum-subtracted energy along the null direction turns negative somewhere along the
geodesic, which is exactly what \eqref{e:smeared-pos}/\eqref{e:cyl-smeared} is meant
to control. Neither example enters that regime. A general proof covering it is
left for future work. 

\section{Possible Implications in Semiclassical Gravity} \label{sec:discussion}
The QFC (and its restricted version) are limited by the first order correction in $G\hbar$. Given its conjectural nature, compared to the proof of the QNEC and INEC in Minkowski space-time, it makes sense to turn the logic upside-down. Assuming that the INEC or QNEC always hold, what is the most general inequality that we can write in the presence of gravity, maintaining the original definitions of generalized entropy $S_{gen}$, quantum expansion $\Theta$ etc.
Consider again the quantum expansion and its derivative: %
\bea 
\label{e:thetat}
\Theta &=& \theta + \frac{4G\hbar}{\mathcal{A}}S^\prime_{\rm out}\,,\\
\label{e:61}
\Theta^\prime &=& \theta^\prime + \frac{4G\hbar}{\mathcal{A}}(S^{\prime \prime}_{\rm out} - \theta S^\prime_{\rm out})\,.
\eea
Substituting the classical Raychaudhuri equation, we get
\bea
\label{e:62}
\Theta^\prime &=& -\frac{1}{d-2}\theta^2-8\pi GT_{\lambda \lambda} + \frac{4G\hbar}{\mathcal{A}}(S^{\prime \prime}_{\rm out} - \theta S^\prime_{\rm out})\,.
\eea
Using \Eqn{e:thetat} and replacing $\theta$,
\bea
\label{e:63}
\Theta^\prime &=& -\frac{1}{d-2}\left(\Theta - \frac{4G\hbar}{\mathcal{A}}S^\prime_{\rm out} \right)^2-8\pi GT_{\lambda \lambda} + \frac{4G\hbar}{\mathcal{A}}\left\{S^{\prime \prime}_{\rm out} - \left(\Theta - \frac{4G\hbar}{\mathcal{A}}S^\prime_{\rm out} \right) S^\prime_{\rm out}\right\}\,,\cr
\Theta^\prime &=& -\frac{1}{d-2}\left\{ \Theta^2 + \left(\frac{4G\hbar}{\mathcal{A}}S^\prime_{\rm out}\right)^2 -\frac{8G\hbar }{\mathcal{A}} S^\prime_{\rm out}\Theta\right\}    -8\pi GT_{\lambda \lambda}+ \frac{4G\hbar}{\mathcal{A}}S^{\prime \prime}_{\rm out} \nonumber \\
&&- \frac{4G\hbar}{\mathcal{A}}\left\{\Theta S^\prime_{\rm out}- \frac{4G\hbar}{\mathcal{A}}(S^\prime_{\rm out})^2 \right\}\,.
\eea
Reshuffling the terms we can write the QNEC as 
\bea
0\leq 8\pi GT_{\lambda \lambda}- \frac{4G\hbar}{\mathcal{A}}S^{\prime \prime}_{\rm out}&=&-\frac{1}{d-2}\left\{ \Theta^2 + \left(\frac{4G\hbar}{\mathcal{A}}S^\prime_{\rm out}\right)^2 -\frac{8G\hbar }{\mathcal{A}}\Theta S^\prime_{\rm out} \right\} \nonumber \\&&- \frac{4G\hbar}{\mathcal{A}}\left\{\Theta S^\prime_{\rm out}- \frac{4G\hbar}{\mathcal{A}}(S^\prime_{\rm out})^2 \right\} - \Theta^\prime \,,
\eea
Since the QNEC has been proven, the right hand side of the equality also has to be non-negative, hence we can get a generalized QFC:
\bea\label{e:65}
\Theta^\prime &\leq& -\frac{1}{d-2}\left\{\Theta^2-(d-3)\left(\frac{4G\hbar }{\mathcal{A}}S^\prime_{\rm out}\right)^2 +\left(d-4\right)\frac{4G\hbar }{\mathcal{A}}\Theta S^\prime_{\rm out}\right\} \,.
\eea
The generalized QFC is not strictly non-positive. It has several interesting features. First, this form of the QFC is closer in spirit to the classical Raychaudhuri equation of $\theta'\leq -\theta^2/(d-2)+\cdots$. Second, all terms involve second order quantum terms. Third, we see that at $d=4$ the violation of non-positivity is strictly of higher order $(G\hbar)^2$:
\be
\Theta^\prime \leq -\frac{1}{2}\left\{\Theta^2-\left(\frac{4G\hbar }{\mathcal{A}}S^\prime_{\rm out}\right)^2\right\}\,.
\ee
This is in line with the original QFC which was a statement only up to order $G\hbar$. Fourth, for $d=3$ both surviving terms are proportional to $\Theta$, and thus give a functional form of the restricted QFC, $\Theta'=\Theta=0$. This will be a manifestation of the fact that at $d=3$, the QNEC and INEC are identical, see below.
Performing a similar exercise for the INEC we have a somewhat different generalized QFC,
\bea
0&\leq& 8\pi GT_{\lambda \lambda}- \frac{4G\hbar}{\mathcal{A}}S^{\prime \prime}_{\rm out}+\frac{d-3}{d-2}\theta \frac{4G\hbar}{\mathcal{A}}S^{\prime}_{\rm out}=-\frac{1}{d-2}\theta \Theta-\Theta^\prime \,,\\
\Rightarrow \Theta^\prime &\leq& -\frac{1}{d-2}\theta \Theta= -\frac{1}{d-2}\left\{\Theta^2-\frac{4G\hbar}{\mathcal{A}}\Theta S^{\prime}_{\rm out}\right\}\,.
\eea
This coincides exactly with the above derivation for $d=3$.
This generalized version gives a functional form of the restricted QFC always, and the only dependence on dimension is in the overall coefficient. Both constructions can serve as useful guides for modified quantum focusing inequalities, that also include higher orders of $G\hbar$.
\section{Conclusions}
\label{sec:conclusions}

We have analyzed the Improved Quantum Null Energy Condition (INEC), obtained from the restricted quantum focusing conjecture. It strengthens
the QNEC by admitting the possibility of non-vanishing
classical expansion. We treated the INEC as a statement in quantum field theory and used
the modular-Hamiltonian techniques for null cuts of the light cone to recast it as the integrated inequality \eqref{e:con1}. Through the convexity of relative entropy \eqref{e:srerel}, the integral is bounded from 
below by a negative semi-definite term, so the INEC holds whenever the smeared null energy integral term dominates it.

We examine this inequality at leading-order
(first-law) limit. For a one-parameter family of states, the relative entropy is of second order, and the condition reduces to the weighted half-line positivity \eqref{e:smeared-pos} of the null energy (Proposition~1). The underlying single-segment operator has a hard endpoint at the cut and it diverges for $d>2$, and is unbounded  from below. Therefore, on the light cone this positivity cannot be guaranteed.

This obstruction is geometric, and it is removed on the Lorentzian
cylinder $\mathbb{R}\times S^{d-1}$. There, the relevant averaged null energy is the
full-geodesic operator \eqref{e:cyl-operator}, whose weight $\sin^d\!\beta$ vanishes
smoothly at both antipodal endpoints; it is well-defined and bounded. Here, the INEC reduces to the
positivity of the partial moment $m_2(\bar A)$. One example where this occurs is when the thermal entropy density dominates. Another example is the thermal (Gibbs) state on the
cylinder, and for this case the INEC holds at leading order with no further assumption (Corollary~1).

Several directions remain open. Beyond leading order, the INEC no longer reduces to a statement about the null energy alone: the relative-entropy convexity \eqref{e:srerel}
enters at second order, and on the cylinder it maps to
$\delta^2 S_{\rm rel}/\delta\bar A^2+2\cot\bar A\,\delta S_{\rm rel}/\delta\bar A\ge0$,
under which the full INEC for a sub-hemisphere cap is governed by the monotonicity of
the vacuum-subtracted cap entropy, $\delta\Delta S_{\rm out}/\delta\bar A\ge0$.
Establishing this monotonicity for a generic state would extend the result beyond the first-law
regime. A fully state-independent proof of the INEC, on the cylinder or directly on the
light cone with a suitably smeared operator, together with a sharper understanding of
the generalized focusing inequalities, are natural next steps.
\section*{ACKNOWLEDGMENTS}
We thank Merav Hadad and Udaykrishna Thattarampilly for collaboration at the early stages of the work. We also thank Viktor Franken, and Francois Rondeau for useful discussions. I.B.D. and A.S. are supported in part by the ``Program
of Support of High Energy Physics'' Grant by the Israeli Council for Higher Education. A.S. acknowledges support from the AGASS center at Ariel University.
\bibliographystyle{JHEP}
\bibliography{inec.bib}

@article{Hollands:2025glm,
    author = "Hollands, Stefan and Longo, Roberto",
    title = "{A New Proof of the QNEC}",
    eprint = "2503.04651",
    archivePrefix = "arXiv",
    primaryClass = "hep-th",
    doi = "10.1007/s00220-025-05450-y",
    journal = "Commun. Math. Phys.",
    volume = "406",
    number = "11",
    pages = "269",
    year = "2025"
}

@article{bousso2022islands,
  title={Islands in closed and open universes},
  author={Bousso, Raphael and Wildenhain, Elizabeth},
  journal={Physical Review D},
  volume={105},
  number={8},
  pages={086012},
  year={2022},
  publisher={APS}
}

@article{engelhardt2021world,
  title={A world without pythons would be so simple},
  author={Engelhardt, Netta and Penington, Geoff and Shahbazi-Moghaddam, Arvin},
  journal={Classical and Quantum Gravity},
  volume={38},
  number={23},
  pages={234001},
  year={2021},
  publisher={IOP Publishing}
}

@article{akers2020quantum,
  title={Quantum maximin surfaces},
  author={Akers, Chris and Engelhardt, Netta and Penington, Geoff and Usatyuk, Mykhaylo},
  journal={Journal of High Energy Physics},
  volume={2020},
  number={8},
  pages={1--43},
  year={2020},
  publisher={Springer}
}

@article{ben2025islands,
  title={Islands in Bianchi type I universe},
  author={Ben-Dayan, Ido and Hadad, Merav and Srivastava, Ayushi},
  journal={Physical Review D},
  volume={111},
  number={4},
  pages={046015},
  year={2025},
  publisher={APS}
}

@article{ben2023islands,
  title={Islands in the fluid: islands are common in cosmology},
  author={Ben-Dayan, Ido and Hadad, Merav and Wildenhain, Elizabeth},
  journal={Journal of High Energy Physics},
  volume={2023},
  number={3},
  pages={1--25},
  year={2023},
  publisher={Springer}
}

@article{hartman2020islands,
  title={Islands in cosmology},
  author={Hartman, Thomas and Jiang, Yikun and Shaghoulian, Edgar},
  journal={Journal of High Energy Physics},
  volume={2020},
  number={11},
  pages={1--56},
  year={2020},
  publisher={Springer}
}

@article{panebianco2021loop,
  title={Loop groups and QNEC},
  author={Panebianco, Lorenzo},
  journal={Communications in Mathematical Physics},
  volume={387},
  number={1},
  pages={397--426},
  year={2021},
  publisher={Springer}
}

@article{Shahbazi-Moghaddam:2022hbw,
    author = "Shahbazi-Moghaddam, Arvin",
    title = "{Restricted quantum focusing}",
    eprint = "2212.03881",
    archivePrefix = "arXiv",
    primaryClass = "hep-th",
    doi = "10.1103/PhysRevD.109.066023",
    journal = "Phys. Rev. D",
    volume = "109",
    number = "6",
    pages = "066023",
    year = "2024"
}

@article{Franken:2025gwr,
    author = "Franken, Victor and Kaya, Sami and Rondeau, Fran{\c{c}}ois and Shahbazi-Moghaddam, Arvin and Tran, Patrick",
    title = "{Tests of restricted Quantum Focusing and a new CFT bound}",
    eprint = "2510.13961",
    archivePrefix = "arXiv",
    primaryClass = "hep-th",
    month = "10",
    year = "2025"
}

@article{Epstein:1965zza,
    author = "Epstein, H. and Glaser, V. and Jaffe, A.",
    title = "{Nonpositivity of energy density in Quantized field theories}",
    doi = "10.1007/BF02749799",
    journal = "Nuovo Cim.",
    volume = "36",
    pages = "1016",
    year = "1965"
}

@article{freivogel2018smeared,
  title={The smeared null energy condition},
  author={Freivogel, Ben and Krommydas, Dimitrios},
  journal={Journal of High Energy Physics},
  volume={2018},
  number={12},
  pages={67},
  year={2018},
  publisher={Springer}
}

@article{blanco2013relative,
  title={Relative entropy and holography},
  author={Blanco, David D and Casini, Horacio and Hung, Ling-Yan and Myers, Robert C},
  journal={Journal of High Energy Physics},
  volume={2013},
  number={8},
  pages={1--65},
  year={2013},
  publisher={Springer}
}

@article{Bekenstein:1973ur,
    author = "Bekenstein, Jacob D.",
    title = "{Black holes and entropy}",
    doi = "10.1103/PhysRevD.7.2333",
    journal = "Phys. Rev. D",
    volume = "7",
    pages = "2333--2346",
    year = "1973"
}

@article{dong2018entropy,
  title={Entropy, extremality, Euclidean variations, and the equations of motion},
  author={Dong, Xi and Lewkowycz, Aitor},
  journal={Journal of High Energy Physics},
  volume={2018},
  number={1},
  pages={1--33},
  year={2018},
  publisher={Springer}
}

@article{engelhardt2015quantum,
  title={Quantum extremal surfaces: holographic entanglement entropy beyond the classical regime},
  author={Engelhardt, Netta and Wall, Aron C},
  journal={Journal of High Energy Physics},
  volume={2015},
  number={1},
  pages={1--27},
  year={2015},
  publisher={Springer}
}

@article{Maldacena:1997re,
    author = "Maldacena, Juan Martin",
    title = "{The Large $N$ limit of superconformal field theories and supergravity}",
    eprint = "hep-th/9711200",
    archivePrefix = "arXiv",
    reportNumber = "HUTP-97-A097, HUTP-98-A097",
    doi = "10.4310/ATMP.1998.v2.n2.a1",
    journal = "Adv. Theor. Math. Phys.",
    volume = "2",
    pages = "231--252",
    year = "1998"
}

@article{dong2014holographic,
  title={Holographic entanglement entropy for general higher derivative gravity},
  author={Dong, Xi},
  journal={Journal of High Energy Physics},
  volume={2014},
  number={1},
  pages={1--32},
  year={2014},
  publisher={Springer}
}

@article{Susskind:1994sm,
    author = "Susskind, Leonard and Uglum, John",
    title = "{Black hole entropy in canonical quantum gravity and superstring theory}",
    eprint = "hep-th/9401070",
    archivePrefix = "arXiv",
    reportNumber = "SU-ITP-94-1",
    doi = "10.1103/PhysRevD.50.2700",
    journal = "Phys. Rev. D",
    volume = "50",
    pages = "2700--2711",
    year = "1994"
}

@article{tHooft:1984kcu,
    author = "'t Hooft, Gerard",
    title = "{On the Quantum Structure of a Black Hole}",
    reportNumber = "Print-84-0924 (UTRECHT)",
    doi = "10.1016/0550-3213(85)90418-3",
    journal = "Nucl. Phys. B",
    volume = "256",
    pages = "727--745",
    year = "1985"
}

@article{herzog2013stress,
  title={Stress tensors from trace anomalies in conformal field theories},
  author={Herzog, Christopher P and Huang, Kuo-Wei},
  journal={Physical Review D—Particles, Fields, Gravitation, and Cosmology},
  volume={87},
  number={8},
  pages={081901},
  year={2013},
  publisher={APS}
}

@article{brown1977stress,
  title={Stress tensors and their trace anomalies in conformally flat space-time},
  author={Brown, Lowell S and Cassidy, James P},
  journal={Physical Review D},
  volume={16},
  number={6},
  pages={1712},
  year={1977},
  publisher={APS}
}

@article{bousso2015new,
  title={New area law in general relativity},
  author={Bousso, Raphael and Engelhardt, Netta},
  journal={Physical review letters},
  volume={115},
  number={8},
  pages={081301},
  year={2015},
  publisher={APS}
}

@article{bousso2016quantum,
  title={Quantum focusing conjecture},
  author={Bousso, Raphael and Fisher, Zachary and Leichenauer, Stefan and Wall, Aron C},
  journal={Physical Review D},
  volume={93},
  number={6},
  pages={064044},
  year={2016},
  publisher={APS}
}

@article{casini2017theorem,
  title={The a-theorem and the Markov property of the CFT vacuum},
  author={Casini, Horacio and Teste, Eduardo and Torroba, Gonzalo},
  journal={arXiv preprint arXiv:1704.01870},
  year={2017}
}

@article{Casini:2023kyj,
    author = "Casini, Horacio and Salazar Landea, Ignacio and Torroba, Gonzalo",
    title = "{Irreversibility, QNEC, and defects}",
    eprint = "2303.16935",
    archivePrefix = "arXiv",
    primaryClass = "hep-th",
    doi = "10.1007/JHEP07(2023)004",
    journal = "JHEP",
    volume = "07",
    pages = "004",
    year = "2023"
}

@article{ceyhan2020recovering,
  title={Recovering the QNEC from the ANEC},
  author={Ceyhan, Fikret and Faulkner, Thomas},
  journal={Communications in Mathematical Physics},
  volume={377},
  number={2},
  pages={999--1045},
  year={2020},
  publisher={Springer}
}

@article{hartman2017averaged,
  title={Averaged null energy condition from causality},
  author={Hartman, Thomas and Kundu, Sandipan and Tajdini, Amirhossein},
  journal={Journal of High Energy Physics},
  volume={2017},
  number={7},
  pages={1--30},
  year={2017},
  publisher={Springer}
}

@article{kontou2013averaged,
  title={Averaged null energy condition in a classical curved background},
  author={Kontou, Eleni-Alexandra and Olum, Ken D},
  journal={Physical Review D—Particles, Fields, Gravitation, and Cosmology},
  volume={87},
  number={6},
  pages={064009},
  year={2013},
  publisher={APS}
}

@article{ford1996averaged,
  title={Averaged energy conditions and evaporating black holes},
  author={Ford, Lawrence H and Roman, Thomas A},
  journal={Physical Review D},
  volume={53},
  number={4},
  pages={1988},
  year={1996},
  publisher={APS}
}

@article{wald1991general,
  title={General proof of the averaged null energy condition for a massless scalar field in two-dimensional curved spacetime},
  author={Wald, Robert and Yurtsever, Ulvi},
  journal={Physical Review D},
  volume={44},
  number={2},
  pages={403},
  year={1991},
  publisher={APS}
}

@article{klinkhammer1991averaged,
  title={Averaged energy conditions for free scalar fields in flat spacetime},
  author={Klinkhammer, Gunnar},
  journal={Physical Review D},
  volume={43},
  number={8},
  pages={2542},
  year={1991},
  publisher={APS}
}

@article{yurtsever1990does,
  title={Does quantum field theory enforce the averaged weak energy condition?},
  author={Yurtsever, Ulvi},
  journal={Classical and Quantum Gravity},
  volume={7},
  number={11},
  pages={L251},
  year={1990},
  publisher={IOP Publishing}
}

@article{faulkner2016modular,
  title={Modular Hamiltonians for deformed half-spaces and the averaged null energy condition},
  author={Faulkner, Thomas and Leigh, Robert G and Parrikar, Onkar and Wang, Huajia},
  journal={Journal of High Energy Physics},
  volume={2016},
  number={9},
  pages={1--35},
  year={2016},
  publisher={Springer}
}

@article{roman1986quantum,
  title={Quantum stress-energy tensors and the weak energy condition},
  author={Roman, Thomas A},
  journal={Physical Review D},
  volume={33},
  number={12},
  pages={3526},
  year={1986},
  publisher={APS}
}

@article{yurtsever1995averaged,
  title={Averaged null energy condition and difference inequalities in quantum field theory},
  author={Yurtsever, Ulvi},
  journal={Physical Review D},
  volume={51},
  number={10},
  pages={5797},
  year={1995},
  publisher={APS}
}

@article{kudler2025covariant,
  title={Covariant regulator for entanglement entropy: Proofs of the Bekenstein bound and the quantum null energy condition},
  author={Kudler-Flam, Jonah and Leutheusser, Samuel and Rahman, Adel A and Satishchandran, Gautam and Speranza, Antony J},
  journal={Physical Review D},
  volume={111},
  number={10},
  pages={105001},
  year={2025},
  publisher={APS}
}

@article{ben2024quantum,
  title={The quantum focusing conjecture and the improved energy condition},
  author={Ben-Dayan, Ido},
  journal={Journal of High Energy Physics},
  volume={2024},
  number={2},
  pages={1--9},
  year={2024},
  publisher={Springer}
}

@article{malik2020proof,
  title={Proof of the quantum null energy condition for free fermionic field theories},
  author={Malik, Taha A and Lopez-Mobilia, Rafael},
  journal={Physical Review D},
  volume={101},
  number={6},
  pages={066028},
  year={2020},
  publisher={APS}
}

@article{koeller2016holographic,
  title={Holographic proof of the quantum null energy condition},
  author={Koeller, Jason and Leichenauer, Stefan},
  journal={Physical Review D},
  volume={94},
  number={2},
  pages={024026},
  year={2016},
  publisher={APS}
}

@article{roy2023proof,
  title={Proof of the R{\'e}nyi quantum null energy condition for free fermions},
  author={Roy, Pratik},
  journal={Physical Review D},
  volume={108},
  number={4},
  pages={045010},
  year={2023},
  publisher={APS}
}

@article{moosa2021renyi,
  title={A R{\'e}nyi quantum null energy condition: proof for free field theories},
  author={Moosa, Mudassir and Rath, Pratik and Su, Vincent Paul},
  journal={Journal of High Energy Physics},
  volume={2021},
  number={1},
  pages={1--49},
  year={2021},
  publisher={Springer}
}

@article{balakrishnan2019general,
  title={A general proof of the quantum null energy condition},
  author={Balakrishnan, Srivatsan and Faulkner, Thomas and Khandker, Zuhair U and Wang, Huajia},
  journal={Journal of High Energy Physics},
  volume={2019},
  number={9},
  pages={1--86},
  year={2019},
  publisher={Springer}
}

@article{bousso2016proof,
  title={Proof of the quantum null energy condition},
  author={Bousso, Raphael and Fisher, Zachary and Koeller, Jason and Leichenauer, Stefan and Wall, Aron C},
  journal={Physical Review D},
  volume={93},
  number={2},
  pages={024017},
  year={2016},
  publisher={APS}
}

@article{wall2012proof,
  title={Proof of the generalized second law for rapidly changing fields<? format?> and arbitrary horizon slices},
  author={Wall, Aron C},
  journal={Physical Review D—Particles, Fields, Gravitation, and Cosmology},
  volume={85},
  number={10},
  pages={104049},
  year={2012},
  publisher={APS}
}

@article{casini2018all,
  title={All the entropies on the light-cone},
  author={Casini, Horacio and Test{\'e}, Eduardo and Torroba, Gonzalo},
  journal={Journal of High Energy Physics},
  volume={2018},
  number={5},
  pages={1--43},
  year={2018},
  publisher={Springer}
}

@article{casini2017modular,
  title={Modular Hamiltonians on the null plane and the Markov property of the vacuum state},
  author={Casini, Horacio and Teste, Eduardo and Torroba, Gonzalo},
  journal={Journal of Physics A: Mathematical and Theoretical},
  volume={50},
  number={36},
  pages={364001},
  year={2017},
  publisher={IOP Publishing}
}

@article{rosso2020global,
  title={Global aspects of conformal symmetry and the ANEC in dS and AdS},
  author={Rosso, Felipe},
  journal={Journal of High Energy Physics},
  volume={2020},
  number={3},
  pages={1--56},
  year={2020},
  publisher={Springer}
}

\end{document}